\documentclass{article}%
\usepackage{amsmath}
\usepackage{graphicx}%
\usepackage{amsfonts}%
\usepackage{amssymb}
\begin{document}

\title{A gauge covariant approximation to QED}
\author{Yuichi Hoshino\\Kushiro National College of Technology\\Otanoshike nishi 2-32-1,Kushiro 084 Hokkaido, Japan}
\maketitle

\begin{abstract}
We examine the fermion propagator in the Dyson-Schwinger equation which is
based on spectral representation.Our approximation preservs vertex
Ward-Takahasi Identity.The subtracted dispersion integral smoothes the
imaginary part of the self energy and removes the infrared divergence in three
dimension.As a result infrared behaviour of the propagator near the threshold
is then found to be similar to four dimensional one.There exists analytic
solutions for arbitrary gauge in four dimension.In our model there is a
possibility of dynamical mass generation in which bare mass vanishes in four dimension.

\end{abstract}

\section*{ Introduction{}}

The non-linear integral Dyson-Schwinger equation has been extensively analysed
with a particular vertex ansatz or in the quenched approximation to deal with
the dynamical symmetry breaking in Quantum Field Theory.Sometimes the
approximations made do not satisfy the Ward-Takahashi (W-T) Identity.In any
case one may use the divergence of the axial-vector current to show dynamical
chiral symmetry breaking via the axial Ward identity,
\[
\partial_{\mu x}\langle T(J_{5}^{\mu}(x)\overline{\psi}(y)\gamma_{5}%
\psi(y))\rangle=-2\delta(x-y)\langle\overline{\psi}\psi\rangle
\,\,\mathrm{\ \ \ and}\,\langle\overline{\psi}\psi\rangle=-i\mathrm{tr}%
(S_{F}(x))
\]
in QED$_{4,3}$ and QCD$_{4}$,where the right hand side of the above equation
depends on the dynamics.Thus an effective mass, induced by gauge interaction
of the fermions,with a non vanishing order parameter $\langle\overline{\psi
}\psi\rangle\neq0$ has been found which is similar to the gap equation in
superconductivity[4].This is the familiar scenario of dynamical symmetry
breaking.But there remains ambiguities as they consider only continuum
contributions in Euclidian space of the fermion self energy and the structure
of the propagator is not clear.Thus we have an interest to see what type of
solutions exist in Minkowski space which satisfy the gauge identities.On the
other hand in the ladder approximation and in the Landau gauge,W-T is valid
only at one loop level of perturbation theory. Atkinson and Blatt[2] have
studied the singurality of the propagator after an analytic continuation of
the Dyson-Schwinger equation to Minkowski space;the only physically meaningful
answer for the propagator is a branch point(cut) on the real axis associated
with zero mass photon,but the analytic continuation from Euclidean to
Minkowski space is not unique and the result depends on the way the vertex is
treated;sometimes it leads to a complex singurality[2].Hereafter we confine
ourselves in QED$_{2+1}.$Main difficulty in ($2+1)$-dimension is the infrared
divergences that was first demonstrated and improved by R. Jackiw and
S.Templeton in the case of massless fermions[1].To aviod the infrared
divergences in QED$_{2+1}$,the infrared behaviour was modified[1,2] by
introducing massless fermions into the photon vaccum polarization since it
affects the low energy behaviour of the photon.After that we explicitly shown
the absence of infrared divergences in the lowest approximation[1].

We shall discuss the structure of the fermion propagator in Minkowski space
using a gauge covariant approximation, which obeys the vertex W-T identity.The
gauge covariant approximation,which is based on dispersion relations,leads to
a linear Dyson-Schwinger equation that admits an analytic solution[3].In this
paper,we analyse a lowest order approximation,where we treat the 2-spinor
representation of fermions in (2+1)-dimension and do not introduce massless
sources[1,2].We derive a linear integral equation for the specral function but
which does not admits analytic solution in the Landau gauge.However we derive
the infrared behaviour of the propagator for arbitrary gauges.We first make a
subtraction of the self-energy as usual in the integral equation and suceed in
avoiding an infrared divergences in 3-dimension;threshold behaviour is
modified to smooth one. In our analysis,the structure of the propagator has an
essential singurality at $p^{2}=m^{2}$ in arbitrary gauge.We also examine the
fermion-antifermion pair condensation since there is no chiral symmetry in our
model and find that the vacuum expectation value $\left\langle \overline{\psi
}\psi\right\rangle $ is finite in three dimension if we remove the ultraviolet
divergence.In four dimension our approximation admits analytic solutions for
arbitrary gauge.We discuss the problems of dynamical chiral symmetry breaking
in four dimension regarding the vanishment of the bare mass $m_{0}Z_{2}%
^{-1}=0$.We compare the results in QED$_{3+1}$ with that in QED$_{2+1}$ in the
same way.

\section{Zeroth Gauge approximation}

W-T identities between Green function in gauge theory are well known.Thus,with
photon legs amputated,the first few identities read
\[
k^{\mu}S(p)\Gamma_{\mu}(p,p-k)S(p-k)=S(p-k)-S(p),
\]%
\begin{equation}
k^{\mu}S(p^{\prime})\Gamma_{\nu\mu}(p^{\prime}k^{\prime}%
;pk)S(p)\!=\!S(p^{\prime})\Gamma_{\nu}(p^{\prime},p^{\prime}\!+\!k^{\prime
})S(p^{\prime}\!+\!k^{\prime})\!-\!S(p\!-\!k^{\prime})\Gamma_{\nu
}(p\!-\!k^{\prime},p)S(p),
\end{equation}
where $S$ is a complete electron propagator and $\Gamma$ stands for the fully
amputated connected Green function; coupling constants have been factorized
out of eq(1).In QED the propagators $S$ and $D$ of fermion and photon,and the
vertex part $\Gamma_{\mu}$ play a central role via Dyson-Schwinger equations
\begin{equation}
1=Z_{2}(\gamma\cdot p-m+\delta m)S(p)-ie^{2}Z_{2}\int\frac{d^{n}k}{(2\pi)^{n}%
}S(p)\Gamma_{\mu}(p,p-k)S(p-k)\gamma_{v}D^{\mu\nu}(k).
\end{equation}
This expression is the same with well known form of renormalized
Dyson-Schwinger equation%

\begin{align*}
S^{-1}  &  =Z_{2}(S_{0}^{-1}+\Sigma),\\
Z_{2}^{-1}S_{0}  &  =S(1+\Sigma S_{0}),
\end{align*}%

\begin{equation}
D_{\mu\nu}^{-1}(k)=Z_{3}[k^{2}g_{\mu\nu}\!-\!k_{\mu}k_{\nu}(1\!-\!\eta
^{-1})]+ie^{2}Z_{2}Tr\int\!\frac{d^{n}p}{(2\pi)^{n}}\gamma_{\nu}%
S(p)\Gamma_{\mu}(p,p-k)S(p-k),
\end{equation}%
\begin{equation}
\Gamma_{\mu}(p,p-k)=Z_{2}\gamma_{\mu}-ie^{2}Z_{2}\int\frac{d^{n}p^{\prime}%
}{(2\pi)^{n}}\gamma_{\lambda}S(p^{\prime})\Gamma_{\nu\mu}(p^{\prime}k^{\prime
};pk)D^{\lambda\nu}(k).
\end{equation}
where $\eta$ is a covariant gauge parameter and we are working in $n$
-dimensions at this stage.In the gauge covariant approximation one seeks
solutions to equations (2)-(4) in the form given above.To this end, begin with
the Lehmann-Kallen spectral representation for the fermion propagator in the
form
\begin{align}
S(p)  &  =(\int_{-\infty}^{-m}+\int_{m}^{\infty})\frac{dw\rho(w)}{\gamma\cdot
p-w+i\epsilon\epsilon(w)}\nonumber\\
&  =\int_{m^{2}}^{\infty}\frac{ds(\gamma\cdot p\rho_{1}(s)+m\rho_{2}%
(s))}{p^{2}-s+i\epsilon}%
\end{align}
where $\rho(w)$ is a positive definite distribution in a non-gauge theory,
\[
\rho(w)=\epsilon(w)\rho(w)\quad\mathrm{where}\quad\epsilon(w)=\theta
(w)-\theta(-w).
\]
In the high energy limit $S\Sigma S_{0}\rightarrow0$,thus the relation between
renormalization constant and the spectral function reads
\begin{align}
S_{0}(p)  &  =Z_{2}S(p),\nonumber\\
\frac{Z_{2}^{-1}}{\gamma\cdot p-m_{0}}  &  =\int_{m^{2}}^{\infty
}ds\frac{\gamma\cdot p\rho_{1}(s)+m\rho_{2}(s)}{p^{2}-s}.
\end{align}
Since
\begin{equation}
S(p-k)-S(p)=\int dw\rho(w)\frac{1}{\gamma\cdot p-w}\gamma\cdot k\frac{1}%
{\gamma\cdot(p-k)-w},
\end{equation}
the simplest possible (but by no means unique) solution of (1) is to take
\begin{equation}
S(p^{\prime})\Gamma_{\mu}^{(0)}(p^{\prime},p)S(p)=\int dw\rho(w)\frac{1}%
{\gamma\cdot p^{\prime}-w}\gamma_{\mu}\frac{1}{\gamma\cdot p-w}.
\end{equation}
The above formula represents a bare vertex weighted by a spectral function
$\rho(w)$ for an electron of mass $w$.This form of the vertex includes the
well known soft photon limit%
\begin{equation}
S(p)\Gamma_{\mu}^{(0)}(p,p)S(p)=-\frac{\partial S(p)}{\partial p_{\mu}}.
\end{equation}
If the ansatz for the vertex in equation (2) is used the renormalized
Dyson-Schwinger equation is written in n-dimension as%

\begin{align}
Z_{2}^{-1}  &  =S(p)(\gamma\cdot p-m_{0})-ie^{2}\int\frac{d^{n}k}{(2\pi)^{n}%
}S(p)\Gamma_{\mu}^{(0)}(p,p-k)S(p-k)\gamma_{\nu}D^{\mu\nu}(k)\nonumber\\
&  =(\gamma\cdot p-m_{0})\int\frac{\rho(w)dw}{\gamma\cdot p-w}-ie^{2}%
\int\frac{d^{n}k}{(2\pi)^{n}}dw\rho(w)\frac{1}{\gamma\cdot p-w}\gamma_{\mu
}\nonumber\\
&  \times\frac{1}{\gamma\cdot(p-k)-w}\gamma_{\nu}(g_{\mu\nu}-\frac{k_{\mu
}k_{\nu}}{k^{2}}(1-\eta))\frac{1}{k^{2}}\nonumber\\
&  =\int\frac{\rho(w)dw}{\gamma\cdot p-w}(\gamma\cdot p-m_{0}+\Sigma(p,w)),
\end{align}
where $\Sigma(p,w)$ is obtained from lowest-order self energy of fermion with
mass $w.$Recalling $Z_{2}^{-1}=\int\rho(w)dw,$equation(10) can be written in
the renormalized form%
\begin{align}
Z_{2}^{-1}  &  =\int d\omega\rho(\omega)\frac{(\gamma\cdot p-m_{0}+\delta
m+\Sigma(p,\omega))}{\gamma\cdot p-\omega+i\epsilon\epsilon(\omega
)}\nonumber\\
&  =\int d\omega\rho(\omega)(1+\frac{\omega+m+\Sigma(p,\omega)-\Sigma
(\omega,\omega)}{\gamma\cdot p-\omega+i\epsilon\epsilon(w)}),\nonumber\\
\delta m  &  =m+m_{0}-\Sigma(m,m).
\end{align}
Here we rewrite this equation in the Klein-Gordon form%
\begin{align}
0  &  =\int d\omega\rho(\omega)\frac{[(\gamma\cdot p+\omega)(m+\omega
+\Sigma(p,\omega)](p,\omega)-(p\rightarrow\omega)}{p^{2}-\omega^{2}%
+i\epsilon\epsilon(w)},\\
\Sigma(p,m)  &  =\gamma\cdot p\Sigma_{1}(p,m)+\Sigma_{2}(p,m)\nonumber
\end{align}
Taking imaginary part yields two eqations ;$\gamma\cdot p\rho_{1}%
(\omega)+\omega\rho_{2}(\omega)=(\gamma\cdot p+\omega)\rho(\omega)$
\begin{align}
&  \int d\omega\frac{\rho_{1}(\omega)\Im\lbrack\omega\Sigma_{1}(p,\omega
)+\Sigma_{2}(p,\omega)-(p\rightarrow\omega)]}{p^{2}-\omega^{2}}\nonumber\\
&  =i\pi\int d\omega\epsilon(\omega)\delta(p^{2}-\omega^{2})\rho_{1}%
(\omega)(\omega-m)\nonumber\\
&  =\frac{i\pi}{2|p|}\int d\omega\epsilon(\omega)(\delta(\omega-p)+\delta
(\omega+p))\rho_{1}(\omega)(\omega-m)
\end{align}%
\begin{align}
&  \int d\omega\frac{\rho_{2}(\omega)\Im\lbrack p^{2}\Sigma_{1}(p,\omega
)+\omega\Sigma_{2}(p,\omega)-(p\rightarrow\omega)]}{p^{2}-\omega^{2}%
}\nonumber\\
&  =i\pi\int d\omega\epsilon(\omega)\delta(p^{2}-\omega^{2})\rho_{2}%
(\omega)\omega(\omega-m)\nonumber\\
&  =\frac{i\pi}{2|p|}\int d\omega\epsilon(\omega)(\delta(\omega-p)+\delta
(\omega+p))\rho_{2}(\omega)\omega(\omega-m),
\end{align}
here the real part does not appear thanks to renormalzation of the self energy
$\Sigma(p,\omega)$ within the dispersion integral.This is the main advantage
of our method and gauge technique.In QED$_{3+1}$the threshold behaviour is
different from QED$_{2+1}$ and $\Im\Sigma_{3+1}(w,w)=0.$In fact it is given as%
\[
\Im\Sigma_{3+1}(p,w)=\frac{e^{2}(p^{2}-w^{2})}{16\pi p^{4}}[\eta(p^{2}%
+w^{2})\gamma\cdot p-(\eta+3)wp^{2}]\theta(p^{2}-w^{2}).
\]
In this case there is no need for subtraction of $\Im\Sigma(p,w)$ at $p=w.$But
QED$_{2+1}$is a super renormalzable theory and there are infrared
singularities.In our method renormalization of the self energy removes the
infrared divergences and the subtracted imaginary part of the self energy
smoothes the threshold behaviour as we see in the next section.Infrared
problem for massless fermion was first discussed by Jakiw and Templeton in
connection with the dressed photon[1].In the case of massive fermion the pair
production is supressed at low energy in the case of massive fermion loop is
concerned for photon vacuum polarization,Therefore the bare photon propagator
is dominant in the infrared.

\section{Zeroth Green function}

Now the self-energy of the fermion can be written as $\Sigma(p,m)=\gamma\cdot
p\Sigma_{1}(p,m)+\Sigma_{2}(p,m)$. ($\Sigma_{1}$ and $\Sigma_{2}$ are often
referred to as the vector and scalar parts of the self-energy respectively.)
In (2+1) dimensions to $O(e^{2})$ and in the photon gauge specified by $\eta
$,the self energy of the fermion is written as follows,
\begin{equation}
\Sigma(p,m)=-ie^{2}\int\frac{d^{3}k}{(2\pi)^{3}}\gamma^{\mu}\frac{\gamma
\cdot(p-k)+m}{(p-k)^{2}-m^{2}}\gamma^{\nu}[\frac{g_{\mu\nu}}{k^{2}%
}-\frac{k_{\mu}k_{\nu}}{k^{4}}(1-\eta)].
\end{equation}
It is convieniant to separate $\Sigma_{1}(p,m)$ and $\Sigma_{2}(p,m)$ by
taking traces of the self-energy,%

\[
Tr(\gamma\cdot p\Sigma(p,m)),Tr(\Sigma(p,m)).
\]
In Euclidean space after angular integrations are performed they can be
written in the following form%
\begin{align}
\Sigma_{1}(p,m)  &  =\frac{\eta e^{2}}{4\pi^{2}p^{2}}\int_{0}^{\infty
}dk\frac{k^{2}}{k^{2}+m^{2}}[1-\frac{p^{2}+k^{2}}{2pk}\ln\left|
\frac{p+k}{p-k}\right| \nonumber\\
\Sigma_{2}(p,m)  &  =\frac{(\eta+2)e^{2}}{4\pi^{2}p}\int_{0}^{\infty
}dk\frac{mk}{k^{2}+m^{2}}\ln\left|  \frac{p+k}{p-k}\right|  .
\end{align}
It can be evaluated by using contour integral$,$%
\begin{align}
\Sigma_{1E}(p,m)  &  =1-\frac{\eta e^{2}m}{8\pi p^{2}}(1-\frac{p^{2}-m^{2}%
}{mp}\tan^{-1}(\frac{p}{m}))\nonumber\\
\Sigma_{2E}(p,m)  &  =\frac{(\eta+2)e^{2}}{4\pi}\frac{m}{p}\tan^{-1}%
(\frac{p}{m}).
\end{align}
In Minkowski space after Wick rotation we get%
\begin{align*}
\Sigma_{1}(p,m)  &  =1+\frac{\eta e^{2}m}{8\pi p^{2}}(1-\frac{p^{2}+m^{2}}%
{mp}\tanh^{-1}(\frac{p}{m}))\\
\Sigma_{2}(p,m)  &  =\frac{(\eta+2)e^{2}}{4\pi}\frac{m}{p}\tanh^{-1}%
(\frac{p}{m}).
\end{align*}
Then the imaginary parts are given by%
\[
\Im\tanh^{-1}(\frac{p}{m})=-i\frac{\pi}{2}\epsilon(\frac{p}{m})\theta
(p^{2}-m^{2})
\]%

\begin{align}
\Im\Sigma_{1}(p,m)  &  =\frac{\eta e^{2}}{16}\frac{p^{2}+m^{2}}{p^{3}}%
\epsilon(\frac{p}{m})\theta(p^{2}-m^{2})\nonumber\\
\Im\Sigma_{2}(p,m)  &  =-\frac{2(\eta+2)e^{2}}{16}\frac{m}{p}\epsilon
(\frac{p}{m})\theta(p^{2}-m^{2}).
\end{align}
The final form of the imaginary part of the self energy is presented in the
following form,%
\begin{equation}
\Im\Sigma_{2+1}(p,m)=\frac{e^{2}}{16}[\gamma\cdot p\frac{p^{2}+m^{2}}{p^{3}%
}-\frac{2(\eta+2)m}{p}]\epsilon(\frac{p}{m})\theta(p^{2}-m^{2}).
\end{equation}
If we substitute the above expression into the equation (13),(14) they can be
written to this order,%
\begin{align}
\epsilon(p)(p-m)\rho_{1}(p)  &  =(\int_{m}^{\epsilon(p)p}-\int_{-p\epsilon
(p)}^{-m})\frac{dw\rho_{1}(w)[\Im((w\Sigma_{1}+\Sigma_{2})(p,w)-(p\rightarrow
w)]}{p^{2}-w^{2}}\nonumber\\
&  =\int_{m}^{p}dw[\rho_{1}(w)+\rho_{1}(-w)]K(p,w),\nonumber\\
K(p,w)  &  =\frac{e^{2}}{8\pi}(-\frac{\eta\left|  p\right|  w}{p^{3}%
}+4\frac{\left|  p\right|  }{p(p+w)})
\end{align}%
\begin{align}
\epsilon(p)(p-m)\rho_{2}(p)  &  =(\int_{m}^{\epsilon(p)p}-\int_{-p\epsilon
(p)}^{-m})\frac{dw\rho_{2}(w)[\Im((p^{2}\Sigma_{1}+w\Sigma_{2}%
)(p,w)-(p\rightarrow w)]}{p^{2}-w^{2}}\nonumber\\
&  =\int_{m}^{p}dw[\rho_{2}(w)-\rho_{2}(-w)]L(p,w),\nonumber\\
L(p,w)  &  =\frac{e^{2}}{8\pi}(-\frac{\eta(w-2p)}{pw}+4\frac{w}{p(p+w)}.
\end{align}
In this way the spectral function obeys the more sensible integral equation.We
assume that the spectral function is real. Delbourgo,Waites[1] derived the
equation with gauge technique but not carried out renormalization.In that case
the equation for the spectral functions contais the real part of the self
energy and it means that the spectral function becomes complex function in the
dispersion theory.Further they could not avoid infrared divergences in
quenched case.In our approximation this point is modified to be consitent with
original equation (11).Thanks to subtraction,kernel $K,L$ does not contain
infrared singularities at $p=w$.We see that it is suitable to use the variable
$p$ instead of $p^{2}$ since the coupling constant $e^{2}$ has a dimension of
mass.First we seek the infrared behaviour of the solution near $p=m.$It is
possible to find in any gauge by virtue of $\rho(w)\rightarrow\rho(p)$ on the
r.h.s of (20),(21) in the limit $p\rightarrow m.$Thus the $\rho$ equation
simplifies to%
\begin{equation}
\frac{d}{dp}[(p-m)\rho(p)]=\rho(p)\lim_{p\rightarrow w\rightarrow m}%
\frac{\Im\Sigma(p,w)}{\pi(p^{2}-m^{2})}=\rho(p)\frac{e^{2}}{8\pi m}(2-\eta).
\end{equation}
It follows that near the electron mass shell%
\begin{align}
\frac{d}{d\omega}[(\omega-1)\rho(\omega)]  &  \simeq\chi(2-\eta)\rho
(\omega)\nonumber\\
\rho(\omega)  &  \simeq C(\omega-1)^{-1+\chi(2-\eta)},\chi=\frac{e^{2}}{8\pi
m},\omega=\frac{p}{m}.
\end{align}
And the propagator near the threshold is then found to be%
\begin{equation}
S(p)_{p^{2}\simeq m^{2}}\sim\frac{(\gamma\cdot p+m)}{p^{2}-m^{2}}%
(\frac{2m^{2}}{p^{2}-m^{2}})^{\chi(\eta-2)}.
\end{equation}
This approximation corresponds to Bloch-Nordsieck Model in 4-dimension,where
only the retarded propagator is considered[3].We see the same feature of
QED$_{2+1},$from the threshod behaviour of the propagator.In QED$_{3+1},$the
Yennie gauge $\eta=3$ produces a free particle pole(Abrikosov1956) in the
lowest approximation:%
\begin{equation}
S(p)_{p^{2}\simeq m^{2}}\sim\frac{(\gamma\cdot p+m)}{p^{2}-m^{2}}(\frac{m^{2}%
}{p^{2}-m^{2}})^{\alpha(3-\eta)/2\pi}.
\end{equation}
In four dimension our method also reproduces the same behaviour in the same
approximation we took above as well as gauge technique[3].Now let us return to
equations (20),(21).Unfortnately we cannot solve the equations
analytically.Instead we use inequalities for the integral kernels that admit
us to evaluate two approximate solutions in the Landau gauge.Notice that the
numerator of the propagator; $\gamma\cdot p\rho_{1}(w)+w\rho_{2}(w),\rho
_{1}(w)$ is odd and $\rho_{2}(w)$ is even in $w.$If we use the inequalities
for (20),(21)%
\[
\frac{1}{2p}\leq\frac{1}{w+p}\leq\frac{1}{w+m},
\]
we derive the following equations for positive and negative $\omega=w/m;$%
\begin{align}
(\omega-1)\rho_{1}(\omega)  &  =2\chi\int_{1}^{\omega}d\omega^{\prime
}\frac{\rho_{1}(\omega)+\rho_{1}(-\omega)}{\omega+\omega^{\prime}}%
,\omega>0\nonumber\\
(\omega+1)\rho_{1}(-\omega)  &  =2\chi\int_{1}^{\omega}d\omega^{\prime
}\frac{\rho_{1}(\omega)+\rho_{1}(-\omega)}{\omega+\omega^{\prime}},\omega<0
\end{align}%
\begin{align}
(\omega-1)\rho_{2}(\omega)  &  =2\chi\int_{1}^{\omega}d\omega^{\prime
}\frac{\omega^{\prime}(\rho_{2}(\omega)-\rho_{2}(-\omega))}{\omega
(\omega+\omega^{\prime})},\omega>0\nonumber\\
(\omega+1)\rho_{2}(-\omega)  &  =2\chi\int_{1}^{\omega}d\omega^{\prime
}\frac{\omega^{\prime}(\rho_{2}(\omega)-\rho_{2}(-\omega))}{\omega
(\omega+\omega^{\prime})},\omega<0
\end{align}
Here we define
\begin{align}
\rho_{1}^{w}(\omega)  &  \equiv\rho_{1}(\omega)-\rho_{1}(-\omega)\nonumber\\
\rho_{2}^{w}(\omega)  &  \equiv\rho_{2}(\omega)+\rho_{2}(-\omega).
\end{align}
It is easy to convert them into differential equations with two cases;%

\begin{equation}
\frac{1}{2\omega}\leq\frac{1}{\omega+\omega^{\prime}}\leq\frac{1}{\omega+1}.
\end{equation}
The rhs is an infrared approximation;$\omega^{\prime}=1(p=m).$The solutions
are derived for $\rho^{w}(\omega)$ and they are given%
\begin{align}
\frac{1}{\omega}(\omega+1)^{-1-\chi}(\omega-1)^{-1+\chi}  &  \leq\rho_{1}%
^{w}(\omega)\leq\exp(-\frac{2\chi}{\omega+1})(\omega+1)^{-2-\chi}%
(\omega-1)^{-1+\chi},\\
\frac{1}{\omega^{2}}(\omega+1)^{-1-\chi}(\omega-1)^{-1+\chi}  &  \leq\rho
_{2}^{w}(\omega)\leq\exp(-\frac{2\chi}{\omega+1})(\omega+1)^{-2-\chi}%
(\omega-1)^{-1+\chi}/\omega.
\end{align}
The functions in both sides are normalized by integrals which define the
normalization conditions for $\rho_{1}(\omega)$ and $\rho_{2}(\omega)$ below%
\begin{align}
\int d\omega\rho_{1}^{U}(\omega)  &  =\int_{1}^{\infty}d\omega\exp
(-\frac{2\chi}{\omega+1})(\omega+1)^{-2-\chi}(\omega-1)^{-1+\chi}%
=N(\chi)\nonumber\\
\int d\omega\rho_{1}^{L}(\omega)  &  =\int_{1}^{\infty}d\omega\frac{1}{\omega
}(\omega+1)^{-1-\chi}(\omega-1)^{-1+\chi}\\
&  =\frac{1}{\chi}+\frac{2\pi}{\sin(\chi\pi)}+2\Phi(-1,1,\chi)=\int
d\omega\omega\rho_{2}^{L}(\omega)
\end{align}%
\[
\int d\omega\omega\rho_{2}^{U}(\omega)=N(\chi)
\]
,where $N(\chi)$ is expressed in terms of Whittaker's function and $\Phi$ is a
Lerch's function.%

\begin{align}
N(\chi)  &  =\frac{\chi^{-1-\frac{\chi}{2}}\exp(-\frac{\chi}{2})M(-\frac{\chi
}{2}+1,\frac{\chi}{2}+\frac{1}{2},\chi)}{4\chi(\chi+1)},\\
M_{\kappa,\mu}(z)  &  =\exp(-\frac{z}{2})z^{\frac{1}{2}+\mu}M(\frac{1}{2}%
+\mu-\kappa,1+2\mu,z),
\end{align}%
\begin{equation}
\Phi(z,a,\nu)=\sum_{n=0}^{\infty}\frac{z^{n}}{(\nu+n)^{a}},
\end{equation}
where $M(a,b,z)$ is a Kummer's function%
\begin{equation}
\frac{\Gamma(b-a)\Gamma(a)}{\Gamma(b)}M(a,b,z)=\int_{0}^{1}dt\exp
(zt)t^{a-1}(1-t)^{b-a-1}.
\end{equation}
In order to agree with the free field limit $\rho(\omega)\rightarrow
\delta(\omega-1)$ for small $\chi$ where the condition $\int\rho
(\omega)d\omega=1$ holds,spectral functions are normalized and renormalized
quantities for small coupling are%
\begin{align}
m_{0}Z_{2}^{-1}  &  =\frac{1}{[n]}\lim_{p^{2}\rightarrow\infty}tr(p^{2}%
S(p))=m\int\omega\rho_{2}(\omega)d\omega\rightarrow m\\
Z_{2}^{-1}  &  =\frac{1}{[n]}\lim_{p^{2}\rightarrow\infty\infty}tr(\gamma\cdot
pS(p))=\int\rho_{1}(\omega)d\omega\rightarrow1.
\end{align}
The normalization constants for $\rho_{1}^{U}=\omega\rho_{2}^{U},\rho_{1}%
^{L}=\omega\rho_{2}^{L}$ are $C^{U}=\chi^{2}/M(-\frac{\chi}{2}+1,\frac{\chi
}{2}+\frac{1}{2},\chi),C^{L}=\chi$ respectively.Here we should notice that two
spectral functions are same in the Landau gauge(i.e.$\rho_{1}(\omega
)=\omega\rho_{2}(\omega)).$It is interesting to examine the possible occurence
of confinement and dynamical mass generation in our model.In general $Z_{2}=0$
is a compositeness condition and $m_{0}=0$ is a signpost of dynamical mass
generation.However in our case two conditions are not satisfied.But the
vacuume expectation value $\left\langle \overline{\psi}\psi\right\rangle $ is
a gauge invariant quantity in general.And the trace of the propagator itself
is gauge independent if the approximation made is sufficent[6].The gauge
transformation of the propagator from the Landau gauge to other covariant
$\eta$ gauge in three dimension is determined$[6]$%
\begin{equation}
S(x,\eta)=\exp(-\frac{\eta e^{2}\left|  x\right|  }{8\pi})S(x,0).
\end{equation}
This shows a gauge invariance of the order parameter$\left\langle
\overline{\psi}\psi\right\rangle $ in three dimension.We examine the order
parameter for fermion-antifermion pair condensation in the case of bare mass
and find%
\begin{align}
\left\langle \overline{\psi}\psi\right\rangle  &  =-itrS(x)=2\int\frac{d^{3}%
p}{(2\pi)^{3}}\int_{1}^{\infty}d\omega\frac{m\omega\rho_{2}(\omega)}%
{p^{2}+m^{2}\omega}\\
-\frac{m^{2}}{2\pi}\chi\int_{1}^{\infty}\omega^{2}\rho_{2}^{L}(\omega)d\omega
&  \geqq\left\langle \overline{\psi}\psi\right\rangle \geqq-\frac{m^{2}}{2\pi
}\int_{1}^{\infty}\omega^{2}\rho_{2}^{U}(\omega)d\omega\nonumber\\
-\frac{m^{2}}{4\pi}\chi &  \geqq\left\langle \overline{\psi}\psi\right\rangle
\geqq-\frac{m^{2}}{2\pi}\frac{4\chi^{2}C}{M(-\frac{\chi}{2}+1,\frac{\chi}%
{2}+\frac{1}{2},\chi)},
\end{align}%
\begin{align}
C  &  =\frac{1}{2}\int_{0}^{1}dt\exp(-\chi t)(1-t)^{\chi-1}-\frac{1}{4}%
\int_{0}^{1}dt\exp(-\chi t)(1-t)^{\chi-1}t\nonumber\\
&  =\frac{1}{2}M(1,\chi+1,-\chi)-\frac{1}{4}M(2,\chi+2,-\chi).
\end{align}
Here we used the dimensional reguralization;%
\[
\int\frac{d^{d}k}{(2\pi)^{D}}\frac{1}{(k^{2}+L)^{a}}=\frac{\Gamma
(a-\frac{D}{2})}{(4\pi)^{\frac{D}{2}}\Gamma(a)}L^{\frac{D}{2}-a},
\]
and took into account the most attractive effect in the long distance and
strong dumps at short distance for the spectral function $\rho_{2}^{U}.$If we
remove the linear ultraviolet divergence the order parameter is finite and
vanishes in the weak couling limit.Thus we conclude that fermion-antfermion
pair condensation occurs in our approximation with finite bare mass.It might
be related to some physical phemomena;for example gapless
superconductivity[9].Here we notice the non-linear relation of the bare mass
and coupling constant,%
\begin{equation}
m_{0}Z_{2}^{-1}=\frac{\chi^{-\frac{1}{2}}\exp(-\frac{\chi}{2})m}{\chi
+1}\&(1+\frac{2\pi\chi}{\sin(\pi\chi)}+\chi\Phi(-1,1,\chi)m
\end{equation}
for $\rho_{2}^{U}$ and $\rho_{2}^{L}.$It is compared with the solution of gap
equation in superconductivity;%
\begin{align}
1  &  =\frac{g}{2(2\pi)^{3}}\int d^{3}p(\epsilon^{2}(p)+\left|  M_{d}\right|
^{2})^{-\frac{1}{2}},\nonumber\\
\left|  M_{d}\right|   &  \simeq\omega_{D}\exp(-\frac{2}{g\nu_{F}}),
\end{align}
where $\nu_{F}$ is the energy density of the number of states of elctrons on
the fermi surface and $\omega_{D}$ is the Debye frequency[10].Finally we show
the results of gauge technique,in whch we set $\gamma\cdot p=p$ in the
infrared.In this case the integral equation for one spectral function we
get.The integral kernels are classified into four pieces

$(1)$ $p>0,w>0$%
\begin{align}
K_{1}(p,w)  &  =(p+w)[(p\Sigma_{1}(p,w)+\Sigma_{2}(p,w))-(p\rightarrow
w)]\nonumber\\
&  =\eta\frac{(p^{2}-w^{2})(p-w)}{p^{2}}+4\frac{(p^{2}-w^{2})}{p}%
\end{align}

$(2)$ $p>0,w<0$%
\begin{align}
K_{2}(p,w)  &  =(p-w)[p\Sigma_{1}(p,-w)+-w\Sigma_{2}(p,-w)-(p\rightarrow
w)]\nonumber\\
&  =\eta\frac{(p-w)^{2}(3p+w)}{p^{2}}+4\frac{(p-w)^{2}}{p}%
\end{align}

$(3)$ $p<0,w>0$%
\begin{align}
K_{3}(p,w)  &  =(-p+w)[-p\Sigma_{1}(-p,w)+\Sigma_{2}(-p,w)-(p\rightarrow
w)]\nonumber\\
&  =-\eta\frac{(p-w)^{2}(3p+w)}{p^{2}}-4\frac{(p-w)^{2}}{p}%
\end{align}

$(4)$ $p<0,w<0$%
\begin{align}
K_{4}(p,w)  &  =-(p+w)[(-p\Sigma_{1}(-p,-w)+\Sigma_{2}(-p,w))-(p\rightarrow
w)]\nonumber\\
&  =-\eta\frac{(p-w)^{2}(3p+w)}{p^{2}}-4\frac{(p-w)^{2}}{p}%
\end{align}%
\begin{align}
1)\text{ }p  &  >0,(p-m)\rho(p)=\int_{m}^{p}\frac{K_{1}\rho(w)+K_{2}\rho
(-w)}{p^{2}-w^{2}}dw\nonumber\\
2)\text{ }p  &  <0,-(p+m)\rho(-p)=\int_{m}^{p}\frac{K_{3}\rho(w)+K_{4}%
\rho(-w)}{p^{2}-w^{2}}dw
\end{align}
These equation could not be solved since it is converted into higher order
differntial equation and not tractable.However the infrared behaviour is
easily determined by choosing only positive $p$ in the kernel $K_{1}$ of
equation (1) and set $p=m$ in front of the integral.The results we get%
\begin{align}
(\omega-1)\rho(\omega)  &  =\chi(\frac{\eta+4}{\omega}\int_{1}^{\omega}%
\rho(\omega)d\omega-\frac{\eta}{\omega^{2}}\int_{1}^{\omega}\omega\rho
(\omega)d\omega),\nonumber\\
\frac{d}{d\omega}[\omega(\omega-1)\rho(\omega)]  &  =\chi((\eta+4)\rho
(\omega)-\eta\omega\rho(\omega)),\\
\rho(\omega)  &  \simeq C(\omega-1)^{-1+^{4\chi}}\omega^{-1+4\chi(\eta
-2)},\chi=\frac{e^{2}}{16\pi m}.
\end{align}
More precisely the spectral function near the threshold is given%
\begin{align}
\rho(\omega)  &  \simeq F(2+z,3+z;4+2z+4\chi;\omega)\omega^{z},\nonumber\\
z  &  =Root(Z^{2}+(4\chi+3)Z+(8\chi+2+\eta\chi)=0).
\end{align}
Near $\omega=1$%
\begin{align}
\rho(\omega)  &  \simeq(\omega-1)^{-1+4\chi}F(2+z,+4\chi,1+z+4\chi
;8+4z+4\chi;1-\omega_{),}\nonumber\\
S(p)_{p^{2}\simeq m^{2}}  &  \sim\frac{\gamma\cdot p+m}{p^{2}-m^{2}%
}(\frac{p^{2}-m^{2}}{2m^{2}})^{-1+4\chi}.
\end{align}
This shows the gauge invariance near the threshold.If we evaluate the spectral
function in perturbative way,first we put $\rho^{(0)}=\delta(w-m), $gauge
dependent part of $\rho^{(1)}$ does not contain $1/(p-m)$ term and does not
contribute the singurality near the threshold.

\section{Analysis in four dimension}

In this section we examine the QED$_{3+1}$ which is familiar but not known
except for the Landau gauge in gauge technique.In our method we show the
analytic solution for the spectral functions in arbitrary gauge.The self
energy of the fermion to $O(e^{2})$ is well known%
\begin{align}
\Sigma(p,m)  &  =\gamma\cdot p\Sigma_{1}(p,m)+\Sigma_{2}(p,m)\nonumber\\
\pi^{-1}\Im\Sigma_{1}  &  =\chi\eta\frac{p^{4}-m^{4}}{p^{4}}\theta(p^{2}%
-m^{2})\nonumber\\
\pi^{-1}\Im\Sigma_{2}  &  =-\chi(\eta+3)m\frac{p^{2}-m^{2}}{p^{2}}\theta
(p^{2}-m^{2}),\chi=\frac{e^{2}}{16\pi^{2}}.
\end{align}
In the same way in the previous section we derive the equation for the
spectral functions%
\begin{align}
(\gamma\cdot p+m)\Im(\gamma\cdot p\Sigma_{1}+\Sigma_{2})  &  =\Im(p^{2}%
\Sigma_{1}+m\Sigma_{2}+\gamma\cdot p(m\Sigma_{1}+\Sigma_{2})),\nonumber\\
\epsilon(p)\frac{p-m}{\left|  p\right|  }\rho_{1}(p)  &  =\frac{2}{\pi}%
\int\rho_{1}(w)dw\frac{\Im(w\Sigma_{1}(p,w)+\Sigma_{2}(p,w))}{p^{2}%
-w^{2}+i\epsilon\epsilon(w)}\nonumber\\
&  =2\chi(\int_{m}^{\epsilon(p)p}-\int_{-\epsilon(p)p}^{-m})(\frac{\eta w^{3}%
}{p^{4}}-\frac{3w}{p^{2}})\rho_{1}(w)dw\\
\epsilon(p)(p-m)\rho_{2}(p)  &  =\frac{2}{\pi}\int\rho_{2}(w)dw\frac{\Im
(p^{2}\Sigma_{1}(p,w)+w\Sigma_{2}(p,w))}{p^{2}-w^{2}+i\epsilon\epsilon
(w)}\nonumber\\
&  =2\chi(\int_{m}^{\epsilon(p)p}-\int_{-\epsilon(p)p}^{-m})(\eta
-\frac{3w^{2}}{p^{2}})\rho_{2}(w)dw
\end{align}
It is helpful to separate $\rho$ into odd and even part in $p;\rho_{2}%
=p\rho_{o}(p^{2})+m\rho_{e}(p^{2})$ then the lhs is%
\[
\epsilon(p)[(p^{2}\rho_{o}-m^{2}\rho_{e})+mp(\rho_{o}-\rho_{e})].
\]
The rhs is even in $p$ then it simplifies to%
\[
p^{2}\rho_{o}-m^{2}\rho_{e}=0\rightarrow\rho_{e}=\frac{p^{2}}{m^{2}}\rho_{o},
\]%
\[
\epsilon(p)mp(\rho_{e}-\rho_{o})=2\chi(\eta\int(w\rho_{o}+m\rho_{e}%
)dw-\frac{3}{p^{2}}\int w^{2}(w\rho_{o}+m\rho_{e})dw.
\]
Introducing dimensional variable $\omega=w/m$%
\begin{align}
\epsilon(\omega)\omega(\omega^{2}-1)\rho_{o}(\omega)  &  =2\chi(\eta\int
\omega^{\prime}(1+\omega^{\prime})\rho_{o}(\omega^{\prime})-\frac{3}%
{\omega^{2}}\int\omega^{\prime3}(\omega^{\prime}+1)\rho_{o}(\omega^{\prime
})d\omega^{\prime})\nonumber\\
(\omega^{2}-1)\rho_{2}(\omega)  &  =2\chi(\eta\int\omega^{\prime}\rho
_{2}(\omega^{\prime})d\omega^{\prime}-\frac{3}{\omega^{2}}\int\omega^{\prime
3}\rho_{2}(\omega^{\prime})d\omega^{\prime}%
\end{align}
where%
\begin{align}
\omega\rho_{o}(\omega^{2})  &  =\rho_{2}(\omega^{2}),s=\omega^{2}\nonumber\\
&  (s-1)\rho_{2}(s)=2\chi(\eta\int\rho_{2}(s^{^{\prime}})ds^{\prime}%
-\frac{3}{s}\int s^{^{\prime}}\rho_{2}(s^{\prime})ds^{\prime},\\
\frac{d^{2}}{ds^{2}}[s(s-1)\rho_{2}(s)]  &  =2\chi((\eta-3)\frac{d}{ds}%
[s\rho_{2}(s)]+\eta\rho_{2}(s)),\\
\rho_{2}(s)  &  =\epsilon(\omega)F(\alpha,\beta;2;s)\omega,\nonumber\\
(\alpha,\beta)  &  =(\frac{3}{2}-\chi(\eta-3)\mp\frac{r^{\prime}}%
{2}),\nonumber\\
r^{^{\prime}2}  &  =(1+6\chi)^{2}+4\chi^{2}\eta^{2}-24\chi^{2}\eta+4\chi\eta.
\end{align}
Next we consider $\rho_{1}=p\rho_{o}+m\rho_{e}.$In the same way to $\rho_{2}$
the equation simplfies to%

\begin{align*}
\epsilon(p)\frac{p-m}{\left|  p\right|  }\rho_{1}(p) &  =2\chi(\frac{\eta
}{p^{4}}\int w^{3}\rho_{1}(w)dw-\frac{3}{p^{2}}\int w\rho_{1}(w)dw),\\
\epsilon(\omega)\frac{\omega}{\left|  \omega\right|  }(\omega^{2}-1)\rho
_{o}(\omega) &  =2\chi(\frac{\eta}{\omega^{4}}\int\omega^{\prime4}%
(\omega^{\prime}+1)\rho_{o}(\omega^{^{\prime}2})d\omega^{\prime}%
-\frac{3}{\omega^{2}}\int\omega^{\prime2}(\omega^{\prime}+1)\rho_{o}%
(\omega^{\prime})d\omega^{\prime}\\
(\omega^{2}-1)\rho_{o}(\omega^{2}) &  =2\chi(\frac{\eta}{\omega^{4}}\int
\omega^{^{\prime}5}\rho_{o}(\omega^{\prime2})d\omega^{\prime}-\frac{3}%
{\omega^{2}}\int\omega^{^{\prime}3}\rho_{o}(\omega^{^{\prime}2})d\omega
^{\prime}.
\end{align*}
The equation can be converted into differential equatison and its solution is
given%
\begin{align}
(s-1)\rho_{o}(s) &  =2\chi(\frac{\eta}{s^{2}}\int s^{\prime2}\rho
_{o}(s^{\prime})ds^{\prime}-\frac{3}{s}\int s^{\prime}\rho_{o}(s^{\prime
})ds^{\prime}),\\
\frac{d^{2}}{ds^{2}}[s^{2}(s-1)\rho_{o}(s)] &  =2\chi((\eta-3)\frac{d}%
{ds}[s^{2}\rho_{o}(s)]-3s\rho_{o}(s)),\\
\rho_{1}(s) &  =\epsilon(\omega)\omega\rho_{o}(s)=\epsilon(\omega
)\frac{F(a,b;2;s)\omega}{s(s-1)^{1+2\chi(3-\eta)}},\nonumber\\
(a,b) &  =(\frac{1}{2}+\chi(\eta-3)\pm\frac{r}{2}),\nonumber\\
r^{2} &  =(1-6\chi)^{2}+4\chi^{2}\eta^{2}-24\chi^{2}\eta-4\chi\eta.
\end{align}
Same to three dimension we get $\rho_{1}(s)=\rho_{2}(s)$ in the Landau
gauge.We obtain the full propagator in the dispersion integral%
\begin{equation}
S_{\eta}(p)=\int_{1}^{\infty}ds\frac{\gamma\cdot p\rho_{1}(s,\eta)+m\rho
_{2}(s,\eta)}{p^{2}-m^{2}s}%
\end{equation}%
\begin{equation}
S(p)=\frac{RD\gamma\cdot p}{p^{2}}(I(a,b;c;1;0)-I(a,b;c;1;\frac{p^{2}}{m^{2}%
})-\frac{RE}{m}R_{2}EI(2-\alpha,2-\beta;c;1;\frac{p^{2}}{m^{2}}),
\end{equation}
where%
\[
c=2\chi(\eta-3),D=\frac{\Gamma(2-a-b)}{\Gamma(2-a)\Gamma(2-b)},E=\frac{\Gamma
(\alpha+\beta-2)}{\Gamma(\alpha)\Gamma(\beta)}.
\]
We have used the basic integral%
\begin{align}
I(a,b;c;d;y) &  =\int_{0}^{\infty}x^{c-1}(x+y)^{-d}F(a,b;c;-x)dx\nonumber\\
&  =\frac{\Gamma(c)\Gamma(a-c+d)\Gamma(b-c+d)}{\Gamma(a+b-c+d)\Gamma
(d)}\nonumber\\
&  \times F(a-c+d,b-c+d;a+b-c+d;1-y).
\end{align}
Here we study the weak coupling limit of the propagator as we formally define
the free field limit.First the vector part we obtain%
\begin{align}
S(p)_{V} &  =\int_{1}^{\infty}ds[\frac{F(a,b;c;1-s)}{(p^{2}-m^{2}%
s)(s-1)^{1-c}}-(p^{2}\rightarrow0)]\frac{\Gamma(2-a-b)}{\Gamma(2-a)\Gamma
(2-b)}\nonumber\\
&  =-\frac{1}{p^{2}}\int_{0}^{\infty}dxx^{c-1}[(x+1-\frac{p^{2}}{m^{2}}%
)^{-1}F(a,b;c;-x)-(p^{2}\rightarrow0)]\frac{\Gamma(2-a-b)}{\Gamma
(2-a)\Gamma(2-b)}\nonumber\\
&  \simeq\frac{1}{p^{2}}\frac{\Gamma(2\chi(\eta-3))\Gamma(1-2\chi(\eta
-3))}{\Gamma(2)}[1-F(2-2\chi\eta,1+6\chi;2;\frac{p^{2}}{m^{2}})]\nonumber\\
&  =\frac{1}{p^{2}}\Gamma(2\chi(\eta-3))\Gamma(1-2\chi(\eta
-3))[1-F(2,1;2;\frac{p^{2}}{m^{2}})]\nonumber\\
&  =\frac{\Gamma(2\chi(\eta-3))\Gamma(1-2\chi(\eta-3))}{p^{2}-m^{2}}.
\end{align}
The scalar part is%
\begin{align}
S(p)_{S} &  =\int_{1}^{\infty}ds\frac{mF(2-\alpha,2-\beta;c;1-s)}{(p^{2}%
-m^{2}s)(s-1)^{1+2\chi(3-\eta)}}\frac{\Gamma(\alpha+\beta-2)}{\Gamma
(\alpha)\Gamma(\beta)}\nonumber\\
&  =\frac{-m}{m^{2}}\int_{0}^{\infty}dxx^{c-1}(x+1-\frac{p^{2}}{m^{2}}%
)^{-1}F(2-\alpha,2-\beta;c;-x)\frac{\Gamma(\alpha+\beta-2)}{\Gamma
(\alpha)\Gamma(\beta)}\nonumber\\
&  \simeq\frac{1}{m}\Gamma(2\chi(\eta-3))\Gamma(1-2\chi(\eta-3))F(2+6\chi
,1-2\chi\eta;2;\frac{p^{2}}{m^{2}})\nonumber\\
&  \rightarrow\frac{\Gamma(2\chi(\eta-3))\Gamma(1-2\chi(\eta-3))}{p^{2}-m^{2}%
},
\end{align}
for weak coupling limit and it behaves as the free propagator.Next we examine
the renormalization constants in this approximation by going to asymptotic
values of $p$ in(68),(69) or else from the formal expression for the%
\begin{align}
Z_{2}^{-1} &  =\int\rho_{1}(s)ds=R\frac{\Gamma(2-a-b)}{\Gamma(2-a)\Gamma
(2-b)}\int_{1}^{\infty}\frac{F(a,b;c;1-s)}{s(s-1)^{-c+1}}ds\nonumber\\
&  =R\Gamma(2\chi(\eta-3))\Gamma(1-2\chi(\eta-3))\rightarrow1
\end{align}
for small $\chi,$then we choose the normalization constant as $R=2\chi
(\eta-3),$and%
\begin{align}
m_{0}Z_{2}^{-1} &  =m\int\rho_{2}(s)ds=mR\frac{\Gamma(\alpha+\beta-2)}%
{\Gamma(\alpha)\Gamma(\beta)}\int_{1}^{\infty}\frac{F(2-\alpha,2-\beta
;c;1-s)}{(s-1)^{c-1}}ds\nonumber\\
&  =\lim_{d\rightarrow0}I(2-\alpha,2-\beta;c;d;y)\rightarrow0.
\end{align}
It is not clear that the use of the formula (67) with limitting case $d=0$ for
$m_{0}$ is correct or not[8].In the high energy limit propagator behaves as a
free one in the weak coupling,we determine two quantities from the $S(p)_{V}$
and $S(p)_{S}.$Then it reads%
\begin{equation}
S(p)_{V}=\frac{\Gamma(1+2\chi(\eta-3))\Gamma(1-2\chi(\eta-3))}{p^{2}%
}[1-F(2-2\chi\eta,1+6\chi;2;\frac{p^{2}}{m^{2}})].
\end{equation}
This shows the residue of inverse power of $p^{2}$ up to normalization
constant $R$%
\begin{equation}
Z_{2}^{-1}=\Gamma(1+2\chi(\eta-3))\Gamma(1-2\chi(\eta-3)).
\end{equation}%
\begin{equation}
S(p)_{S}=\frac{\Gamma(1+2\chi(\eta-3))\Gamma(1-2\chi(\eta-3))}{m}%
F(2+6\chi,1-2\chi\eta;2;\frac{p^{2}}{m^{2}}).
\end{equation}
Let us expand this in inverse powers of $p^{2}.$The results is
\begin{align}
S(p)_{S} &  =\frac{\Gamma(1+2\chi(\eta-3))\Gamma(1-2\chi(\eta-3))}%
{m}\nonumber\\
&  \times\lbrack\frac{\Gamma(1+2\chi(\eta+3))}{\Gamma(2+6\chi)\Gamma
(1+2\chi\eta)}(-\frac{m^{2}}{p^{2}})^{1-2\chi\eta}F(1-2\chi\eta,-2\chi
\eta;2+2\chi(\eta+3);\frac{m^{2}}{p^{2}})\nonumber\\
&  +\frac{\Gamma(-1-2\chi(\eta+3))}{\Gamma(1-2\chi\eta)\Gamma(-6\chi
)}(-\frac{m^{2}}{p^{2}})^{2+6\chi}F(2+6\chi,1+6\chi;-2\chi(\eta+3);\frac{m^{2}%
}{p^{2}})].
\end{align}
From this form we find that the first term corresponds to free field%
\begin{align}
\lim_{p^{2}\rightarrow\infty}[p^{2}S(p)_{S}] &  =m_{0}Z_{2}^{-1}%
=m\frac{\Gamma(1-2\chi(\eta-3))\Gamma(1+2\chi(\eta-3))\Gamma(1+2\chi(\eta
+3))}{\Gamma(1+2\chi\eta)\Gamma(2+6\chi)}\nonumber\\
&  \times\left[
\begin{array}
[c]{cc}%
1 & \eta=0\\
0 & \eta<0\\
\infty & \eta>0
\end{array}
\right]  ,
\end{align}
for small $\chi.$The second case in the parenthesis corresponds to dynamical
mass generation.In Ref[8],the relation between the vanishment of the bare
fermion mass and dynamical chiral symmetry breaking was discussed in QED and
QCD.In both cases $m_{0}Z_{2}^{-1}=0$ was explicitly shown using
renormalization group equations for the propagator;%
\begin{align}
m_{0}Z_{2}^{-1} &  =\lim_{p^{2}\rightarrow\infty}p^{2}S(p)_{S}=m\int\rho
_{2}(s)ds=0,\nonumber\\
\frac{m_{0}}{m} &  =\frac{\int\rho_{2}(s)ds}{\int\rho_{1}(s)ds}=0,
\end{align}
where the existence of the ultraviolet stagnant point was assumed in QED;%
\begin{equation}
\lim_{\kappa\rightarrow\infty}g(\kappa)=g_{\infty},\beta(g_{\infty}%
)=0,\beta^{\prime}(g_{\infty})<0.
\end{equation}
We considered the infrared and ultraviolt boundary conditions.The spectral
functions satisfying boundary conditions are
\begin{equation}
\rho_{1}(s)=R\frac{\Gamma(2-a-b)}{\Gamma(2-a)\Gamma(2-b)}\frac{F{(}%
a{,}b{;2\chi(\eta-3);1-}s{)}}{s{(}s{-1)^{1+2\chi(3-\eta)}}},
\end{equation}%
\begin{equation}
\rho_{2}(s)=R\frac{\Gamma(\alpha+\beta-2)}{\Gamma(\alpha)\Gamma(\beta
)}\frac{F(2-\alpha,2-\beta;2\chi(\eta-3);1-s)}{(s-1)^{1+2\chi(3-\eta)}}.
\end{equation}
It is clear that the infrared behaviour is the same as Bloch-Nordsieck Model
and in any gauge the spectral functions are normalizable for weak coupling in
a sense of analytic continuation.Here we show the results in the Gauge
Technique solution in the Landau gauge;%

\begin{equation}
\rho(s)=\frac{2\epsilon R(\epsilon)}{m(s-1)^{1+2\epsilon}}[\frac{1}%
{w}F(-\epsilon,-\epsilon;-2\epsilon;1-s)+F(-\epsilon,1-\epsilon;-2\epsilon
;1-s)]
\end{equation}
with $\epsilon=3e^{2}/16\pi^{2},s=(w/m)^{2}$ and $R(\epsilon)$ is the
normailzation factor[7].Notice that first term corresponds to $\rho_{1}$ (odd
in $w)$ and second term corresponds to $\rho_{2}$(even in $w$ ).Next we
examine the pair condensation.The vacuum expectation value $\left\langle
\overline{\psi}\psi\right\rangle $ is proportional to%
\begin{equation}
-itrS(x)=4\int\frac{d^{4}p}{(2\pi)^{4}}S(p)_{S}=-4C\int\frac{d^{4}p}%
{(2\pi)^{4}}\int_{1}^{\infty}ds\frac{m\rho_{2}(s)}{p^{2}+m^{2}s}.
\end{equation}
In equations (59),(62) spectral functions in the Landau gauge have a simple
form%
\begin{equation}
\rho_{1}(s)=\rho_{2}(s)=\frac{C}{s(s-1)^{1+6\chi}}.
\end{equation}
We change the variable $s\rightarrow1/s,$and get%
\begin{equation}
\rho_{2}(s)=-\frac{6\chi s^{6\chi}}{(1-s)^{1+6\chi}},Z_{2}^{-1}=\int_{0}%
^{1}\rho_{1}(s)ds=\Gamma(1+6\chi)\Gamma(1-6\chi).
\end{equation}
Then we obtain the propagator%
\begin{align}
S(p)  &  =-\frac{1}{m^{2}}\int_{0}^{1}\frac{(\gamma\cdot p+m)s\rho_{2}%
(s)}{sp^{2}-m^{2}}ds\nonumber\\
&  =-\frac{(\gamma\cdot p+m)6\pi\chi(1+6\chi)}{m^{2}\sin(6\pi\chi)}%
F(1,2+6\chi;2;\frac{p^{2}}{m^{2}})\nonumber\\
&  =\frac{\Gamma(1+6\chi)\Gamma(1-6\chi)(\gamma\cdot p+m)}{p^{2}%
}(1-(1-\frac{p^{2}}{m^{2}})^{-1-6\chi}).
\end{align}
In this form bare mass and renormalization constant are
\begin{equation}
m_{0}Z_{2}^{-1}=m\Gamma(1+6\chi)\Gamma(1-6\chi),Z_{2}^{-1}=\Gamma
(1+6\chi)\Gamma(1-6\chi).
\end{equation}
If we use dimensional reguralization%
\begin{equation}
\int\frac{d^{D}p}{(2\pi)^{D}}\frac{1}{(p^{2}+L)^{a}}=\frac{\Gamma
(a-\frac{D}{2})}{(4\pi)^{\frac{D}{2}}\Gamma(a)}L^{\frac{D}{2}-a}%
,D=4-2\epsilon,a=1,
\end{equation}
we can avoid quadratic and infrared divergence and can evaluate the finite
answer%
\begin{align}
\left\langle \overline{\psi}\psi\right\rangle  &  =-itrS(x)=-4\int
\frac{d^{4}p}{(2\pi)^{4}}\int_{1}^{\infty}ds\frac{m\rho_{2}(s)}{p^{2}+m^{2}%
s}\nonumber\\
&  =4\lim_{\epsilon\rightarrow0}\frac{\Gamma(-1+\epsilon)}{16\pi^{2}}\int
_{1}^{\infty}s^{1-\epsilon}m^{3}\rho_{2}(s)ds\nonumber\\
&  =\lim_{\epsilon\rightarrow0}\frac{m^{3}\Gamma(-1+\epsilon)\Gamma
(2+6\chi+\epsilon)\Gamma(1-6\chi)}{4\pi^{2}\Gamma(\epsilon)}=-\frac{m^{3}%
}{4\pi^{2}}\Gamma(2+6\chi)\Gamma(1-6\chi).
\end{align}
In general case if we apply the same formula (67),we obtain%
\begin{align}
&  R\frac{\Gamma(\alpha+\beta-2)}{\Gamma(\alpha)\Gamma(\beta)}\int_{1}%
^{\infty}\frac{s^{1-\epsilon}F(2-\alpha,2-\beta;2\chi(\eta-3);1-s)ds}%
{(s-1)^{1+2\chi(3-\eta)}}\nonumber\\
&  =R\frac{\Gamma(\alpha+\beta-2)}{\Gamma(\alpha)\Gamma(\beta)}\int
_{0}^{\infty}x^{2\chi(\eta-3)-1}(x+1)^{1-\epsilon}F(2-\alpha,2-\beta
;2\chi(\eta-3);-x)dx\nonumber\\
&  =\frac{\Gamma(1+2\chi(\eta-3))\Gamma(6\chi+\epsilon)\Gamma(\epsilon
-1-2\chi\eta)}{\Gamma(\epsilon)\Gamma(\epsilon-1)}.
\end{align}

Thus the naive limit $\epsilon\rightarrow0$ leads $\left\langle \overline
{\psi}\psi\right\rangle =0$ except for $\eta=0.$In the Landau gauge the result
coincides the previous case.This result is suspicious for us too.Therefore we
integrate explicitly in momentum space by changing variable $x\rightarrow-x$
and get%
\begin{equation}
\int_{0}^{\infty}\frac{d^{4}p}{(2\pi)^{4}}S(ip)_{S}\simeq\int_{0}^{\infty
}xF(a,b;2;-x)dx=0,
\end{equation}
here we use the formula%
\begin{equation}
\int_{0}^{\infty}x^{-s-1}F(a,b;c;-x)dx=\frac{\Gamma(a+s)\Gamma(b+s)\Gamma
(c)\Gamma(-s)}{\Gamma(a)(b)\Gamma(c+s)}.
\end{equation}
If $c=2$ and $s=-2$ it yields the same result with (71).Here we summarise our
calculus of the vacuum expectation value.%
\begin{align}
\left\langle \overline{\psi}\psi\right\rangle  &  =finite;m_{0}Z_{2}^{-1}%
\neq0,\eta=0\nonumber\\
&  =0;m_{0}Z_{2}^{-1}=0,\infty,\eta\neq0.
\end{align}
In the usual treatment of dynamical mass generation, if the bare mass vanishes
order parameter becomes finite but not zero.In that case the Dyson-Schwinger
equation is non-linear and non-trivial mass function $(i.e.$%
scalar\ part\ of\ the\ propagator) made it convergent and finite.In our model
non-trivial propagator does not necessary leads to non-vanishing vacuum
expectation value.

\section{Improved Vertex function}

Direct substitution of our solution $\rho(w)$ into equation $(8)$, gives the
zeroth order vertex function,
\begin{equation}
S(p)\Gamma_{\mu}^{(0)}S(p^{\prime})=\int dw\rho(w)\frac{1}{\gamma\cdot
p-w}\gamma_{\mu}\frac{1}{\gamma\cdot p^{^{\prime}}-w}.
\end{equation}
We may improve the vertex by adding transverse combinations of terms such as
\[
\lbrack(p_{\mu}\gamma\cdot p^{^{\prime}}+p_{\mu}^{^{\prime}}\gamma\cdot
p-p\cdot p^{^{\prime}}\gamma_{\mu})+((p+p^{^{\prime}})_{\mu}+i(p^{^{\prime}%
}-p)_{\alpha}\epsilon_{\alpha\mu\nu}\gamma_{\nu})w+w^{2}\gamma_{\mu}]
\]
in the numerator of the integral, allowing for a parity violating
contribution.One may also include a form factor proportional to a linear
combination of $\gamma_{\mu}$ and $(p+p^{^{\prime}})_{\mu}$ and try to ensure
that on-shell quantities are gauge invariant.

\section{Summary}

To solve the Dyson-Schwinger equation in QED in four and three dimension,
gauge technique has been used[1],[4].In the annalysis in four dimension,it
seems correct in the infrared but one is suspicious to other kinematical
region and to its restriction only in the Landau gauge.For example in the
Yennie gauge there is no solution due to the ultraviolet divergences.In three
dimension the difficulty of infrared divergences had been suggested to modify
the photon propagator or the introduction of Chern-Simons term.However the
latter violates parity and is induced by quantum loop correction.Previously,to
soften the infrared divergence,massless fermions were introduced into photon
vacuum polarization[1] but in that case no renormalization of the self energy
was done.The main difficulty in that case is the use of the variable
$\omega^{2}$ for the spectral function $\rho(\omega^{2})$ in spite of the
dimension-full coupling constant $e^{2}$.In this work we have instead analyzed
the quenched case with renormalization that leads to avoid the infrared
singularity.We estimated the solution of the spectral function $\rho(\omega)$
in the Landau gauge; these are simple functions like in $QED_{3+1}[3].$ The
infrared behaviour in the arbitrary gauge is also determined.The gauge
dependence near the mass shell is similar to that in $QED_{3+1}$ thanks to the
renormalization; $\rho(\omega)\simeq(\omega-1)^{-1+\chi(2-\eta)}$.There is
some difference in the infrared behaviour between our method and gauge
technique.The latter show the gauge independence near the threshold.Here we
summarize the differences between two approximations.The former shows the high
energy behaviour of the spectral function is $1/p^{2}$ and the cut structure
near the mass shell is given by massless fermion-loop;$\rho(p)$
$=(p-m)^{-1-e^{2}/c\pi^{2}}$ for $p-m\ll e^{2}$ and $c=e^{2}N/8$ for $N$
massless fermions;the integral equation for $\rho(\omega)$ was not solved
analytically because of its complexity due to higher order corrections.In our
approximation the high energy behaviour is $1/p^{3}$ for $\rho$,and the
structure near the mass shell is determined by coupling constant mass ratio
$\chi=e^{2}/(8\pi m)$.Fermion antifermion pair condensation parameter is gauge
invariant but approximation makes it not.The order parameter is finite if we
remove the linear ultraviolet divergence and vanishes in the weak coupling
limit with finite bare mass in our approximation.Both in three and four
dimension,we find the non-linear relation between bare mass and coupling
constant.They differ from the solution of gap equation in super
conductivity.Since $QED_{2+1}$ is super renormalizable,ordinary
renormalization group is not relevant,but the gauge covariant approximation is
non-perturbative,and we have succeeded in finding the infrared behaviour of
the fermion propagator.In four dimension there exists analytic solutions for
the spectral functions which satisfy the boundary condition for arbitrary
gauge instead gauge technique admits them for the Landau gauge.Both of them
show the well known infrared behaviour as Bloch-Nordsieck.Without use gauge
technique $\gamma\cdot p=p$ , ultraviolet divergence is absent for arbitrary
gauge in the equation for the spectral functions.The vanishment of bare
fermion mass occurs in the negative gauge parameter $\eta$ in our quenched
approximation.The vacuum expectation value $\left\langle \overline{\psi}%
\psi\right\rangle $ is shown to be finite which is different from the quenched
ladder approximation in the Landau gauge[4].

\section{Acknowledgement}

The author is indebted to Prof.Delbourgo during his stay in March 2000 at the
University of Tasmania for his hospitality and stimulating comments.He also
thanks to Dr.Suzuki at Hokkaido University for his careful comments and discussion.

\section{References}

\noindent\lbrack1]R.Jakiw,S.Templeton,Phys.Rev.\textbf{23D}%
.2291(1981);\newline
T.Appelquist,D.Nash,L.C.R.Wijewardhana,Phys.Rev.Lett.\textbf{60}(1988)2575;
\newline A.B.Waites,R.Delbourgo,Int.J.Mod.Phys.\textbf{27A}%
(1992)6857;\newline Y.Hoshino,T.Matsuyama,Phys.Lett.\textbf{222B}(1989)493.

\noindent\lbrack2]D.Atkinson,D.W.E.Blatt,Nucl.Phys.\textbf{151B}%
(1979)342;\newline P.Maris, Phys.Rev.\textbf{52D}%
(1995)6087;Y.Hoshino,NuovoCim.\textbf{112A}(1999)335.

\noindent\lbrack3]F.Bloch,A.Nordsieck,Phys.Rev.\textbf{52(}%
1937)54;Abrikosov,A.A,JETP,\textbf{30},96(1956);\newline
R.Delbourgo,P.West,J.Phys.\textbf{10A}%
(1977)1049;R.Delbourgo,Phys.Lett.\textbf{72B}(1977)96.

\noindent\lbrack4]T.Maskawa,H.Nakajima,Prog.Theor.Phys.\textbf{52}%
(1974)1326,\newline Prog.Theor.Phys.\textbf{54}%
(1975)860;R.Fukuda,T.Kugo,Nucl.Phys.\textbf{B117}(1976)250,\newline
H.D.Politzer,Nucl.Phys.\textbf{B117}(1976)397.

\noindent[5]S.Deser,R.Jackiw,S.Templeton,Ann.Phys.\textbf{140}(1982)372.

\noindent\lbrack6]C.J.Burden,J.Praschifka,C.D.Roberts,Phys.Rev.\textbf{46D}%
(1992)2695;\newline L.D.Landau,I.M.Kharatonikov,Zk.Eksp.Theor.Fiz.\textbf{29}%
.89(1958);\newline B.Zumino,J.Math.Phys.\textbf{1}(1960)1;\newline
Landau,Lifshits''Quantum Electrodynamics'', Butterworth Heinemann.
\hfill\hfill\hfill\newline [7]R.Delbourgo,IL Nuovo Cim.\textbf{49A}(1979)485.
\newline [8]R.Delbourgo,B.W.Keck;J.Phys.G\textbf{6(}{\normalsize 1980)275}%
;\newline K.Nishijima,Prog.Theor.Phys.\textbf{81}(1989)878\newline [9]in
H.Nakajima''Introduction to Superconductivity'',Baihukan in Japanese.
\newline [10]in V A Miransky''Dynamical Symmetry Breaking in Quantum Field
Theories'',World Scien
\end{document}